\documentclass[epj]{svjour}
\usepackage{amssymb}
\usepackage{amsmath}
\usepackage{graphics}

\begin{document}

\title{Impedance measurement technique for quantum systems}
\author{S.N. Shevchenko}
\offprints{}
\institute{B.Verkin Institute for Low Temperature Physics and Engineering, 47 Lenin
Ave., Kharkov 61103, Ukraine.}
\date{Received: date / Revised version: date}

\abstract{The impedance measurement technique consists in that the phase-dependent
(parametric) inductance of the system is probed by the classical tank
circuit via measuring the voltage. The notion of the parametric inductance
for the impedance measurement technique is revisited for the case when a
quantum system is probed. Measurement of the quantum state of the system of
superconducting circuits (qubits) is studied theoretically. It is shown that the result of the
measurement is defined by the partial energy levels populations in the qubits.
 } 

\maketitle

\section{Introduction}

The supercurrent $I$, flowing through a weak link between two bulk
superconductors with the phase difference $\phi $, has the properties as a
nonlinear inductor. This can be described by introducing the phase-dependent
(parametric) inductance $\mathcal{L}=(\Phi _{0}/2\pi )\left( \partial I/\partial
\phi \right) ^{-1}$. If the weak link is included in a ring, then the phase $%
\phi $ is related to the magnetic flux $\Phi $, piercing the ring, $\phi
=2\pi \Phi /\Phi _{0}$. The parametric inductance can be measured \cite%
{SilZim} and being inductively coupled to the resonant $LC$ tank circuit
provides the tool to measure the flux $\Phi $ \cite{Likharev}, \cite%
{Shnyrkov}. The effective inductance of the tank circuit depends on the
parametric inductance $\mathcal{L}(\phi )$ and current $I(\phi )$ and thus the
measurement in the tank circuit can be used for finding the inductance $\mathcal{L}
$ \cite{RandD}, which is the so-called impedance measurement technique.

The impedance measurement technique was recently applied for the measurement
of the small currents in mesoscopic samples \cite{ili} and it was proposed
for the description of the currents in superconducting qubits \cite{Zorin02}%
, \cite{Krech02}, \cite{Greenberg02a}, \cite{Smirnov} and the series of the
experimental results were obtained \cite{Ilichev04}. However the theoretical
works in this field consider mostly the ground state. If the superconducting
qubit is excited to the upper state, then the current in it has the
probabilistic character, and in this way the parametric inductance depends
not only on the clockwise (counter-clockwise) current value but also on the
probabilities of the respective states. This consideration was used for the
description of the phase-biased charge qubit \cite{multiphoton}. And in this
paper we study in detail the specifics of the impedance measurement
technique when the quantum system is probed. The detailed presentation is
aimed to show how the tank circuit is influenced by the parametric
inductance, and how this inductance have to be treated for the system of
coupled superconducting circuits (qubits). For concreteness we consider the
superconducting circuits to be either flux or phase-biased charge qubits.
The flux qubit \cite{Mooij} consists of a loop with three Josephson
junctions. The phase-biased charge qubit \cite{Zorin02}, \cite{Krech02}
consists of a loop with two closely situated Josephson junctions and with
the gate which controls the charge on the island between the junctions.

\section{Tank circuit coupled to quantum object}

\subsection{Equations for tank circuit}

The quantum system (coupled superconducting qubits) is considered to be
weakly coupled via a mutual inductance $M$ to the classical tank circuit.
The circuit consists of the inductor $L_{T}$, capacitor $C_{T}$, and the
resistor $R_{T}$ connected in parallel. The tank circuit is biased by the
current $I_{bias}$, and the voltage on it $V_{T}$ can be measured. To obtain
the equation for the voltage, we write down the system of equations, for the
current in the three branches, namely, through the inductor ($I_{L}$), the
capacitor ($I_{C}$), and the resistor ($I_{R}$) (see \textit{e.g.} in the
Chapter 14 of Ref. \cite{Likharev}):%
\begin{eqnarray}
I_{bias} &=&I_{L}+I_{C}+I_{R},  \label{system} \\
I_{C} &=&\dot{e},\text{ }e=C_{T}V_{T}, \\
I_{R} &=&V_{T}/R_{T}, \\
V_{T} &=&L_{T}\dot{I}_{L}-\dot{\Phi}_{e},  \label{4}
\end{eqnarray}%
where $e$ is the charge at the capacitor plate, the dot stands for the time
derivative, $\Phi _{e}$ is the flux through the tank circuit. This flux is
the response of the quantum system to the flux, induced in it by the current
$I_{L}$, and its time derivative equals (see below for details):%
\begin{equation}
\dot{\Phi}_{e}=\widetilde{L}\dot{I}_{L},  \label{Fi_e}
\end{equation}%
and thus equation (\ref{4}) can be rewritten by introducing the effective
inductance of the tank circuit $L_{eff}$:%
\begin{eqnarray}
V_{T} &=&L_{eff}\dot{I}_{L}, \\
L_{eff} &=&L_{T}-\widetilde{L}.  \label{Leff}
\end{eqnarray}%
Then from the system of equations (\ref{system}-\ref{4}) we derive the
equation for the voltage in the tank circuit:%
\begin{equation}
C_{T}\overset{\centerdot \centerdot }{V}_{T}+R_{T}^{-1}\dot{V}%
_{T}+L_{eff}^{-1}V_{T}=\dot{I}_{bias}.  \label{eq_for_V}
\end{equation}

\subsection{Effective inductance of qubits}

Now we derive the relation (\ref{Fi_e}); consider the flux $\Phi
_{e}=\sum\nolimits_{i}\Phi _{e}^{(i)}$, where $\Phi _{e}^{(i)}$ is the flux
induced by $i$-th qubit in the tank circuit: $\Phi
_{e}^{(i)}=M_{iT}I_{qb}^{(i)}$. Here $M_{iT}$ is the mutual inductance of
the qubit and the circuit, $I_{qb}^{(i)}$ is the current in the $i$-th qubit
which equals to the expectation value of the current operator: $%
I_{qb}^{(i)}=\left\langle \widehat{I}_{i}\right\rangle =Sp\left( \widehat{%
\rho }\widehat{I}_{i}\right) $, where $\widehat{\rho }$ is the reduced
density matrix of the system of qubits. (Note that for one qubit,
substituting $\Phi _{e}=MI_{qb}$, Eq. (\ref{eq_for_V}) coincides with Eq.
(15) in Ref. \cite{Greenberg02a}.)

The total flux that threads the loop of the $i$-th qubit $\Phi ^{(i)}$
consists of the external magnetic flux $\Phi _{x}^{(i)}$ and the
self-induced flux $-L_{i}I_{qb}^{(i)}$ ($L_{i}$ is the geometrical
inductance of the loop):%
\begin{equation}
\Phi ^{(i)}=\Phi _{x}^{(i)}-L_{i}I_{qb}^{(i)}.  \label{fluxon}
\end{equation}%
This equation can be rewritten by introducing the parametric inductance,%
\begin{equation}
\mathcal{L}_{i}^{-1}=\frac{\partial I_{qb}^{(i)}}{\partial \Phi ^{(i)}},
\label{inductance}
\end{equation}%
to relate the variations of the external flux through the qubit $\delta \Phi
_{x}^{(i)}$ and of the current in it $\delta I_{qb}^{(i)}$, as following:%
\begin{equation}
\delta \Phi _{x}^{(i)}=\delta \Phi ^{(i)}+L_{i}\delta I_{qb}^{(i)}=\left(
\mathcal{L}_{i}+L_{i}\right) \delta I_{qb}^{(i)}.
\end{equation}%
Thus, we obtain the variation of the flux induced by the qubit in the tank
circuit:%
\begin{equation}
\delta \Phi _{e}^{(i)}=M_{iT}\delta I_{qb}^{(i)}=\frac{M_{iT}}{\mathcal{L}_{i}+L_{i}}%
\delta \Phi _{x}^{(i)}.  \label{var}
\end{equation}

The flux $\Phi _{x}^{(i)}$ in the $i$-th qubit is considered to consist of
the fluxes induced by the tank circuit, $M_{iT}I_{L}$, by the microwave
source, $\Phi _{ac}^{(i)}\sin \omega t$, and by\ additional lines and by
other qubits, $\Phi _{shift}^{(i)}$ \cite{2qbs}:%
\begin{equation}
\Phi _{x}^{(i)}(t)=M_{iT}I_{L}(t)\,+\Phi _{shift}^{(i)}+\Phi _{ac}^{(i)}\sin
\omega t.  \label{Fi_x}
\end{equation}%
$\,$Now let us recall that we consider the variation of the flux in order to
calculate the derivative in time $\dot{\Phi}_{e}$ which have to be
substituted in Eq. (\ref{4}), which describe the tank circuit. We note that
usually the dynamics of the tank circuit (with the frequency $\omega _{rf}$
close to the resonant frequency $\omega _{T}=\left( L_{T}C_{T}\right) ^{-1/2}
$) is significantly slower than the dynamics of a qubit driven by the
microwave source at frequency $\omega \gg \omega _{T}$. Hence, being
interested in the response of the measurement system (that is of the tank
circuit), we average equations over the period $2\pi /\omega $. After this
averaging the component $\Phi _{ac}^{(i)}\sin \omega t$ tends to zero and we
have: $\delta \Phi _{x}^{(i)}\approx M_{iT}\delta I_{L}$. Here and below the
time-averaging is assumed. (Note that here it is assumed that the
expectation value for the current $I_{qb}^{(i)}$ in Eq. (\ref{var}) weakly
depend on time during the time interval of the order of $2\pi /\omega $ and
its time dependence is defined by the tank circuit dynamics only.) Then it
follows%
\begin{equation}
\dot{\Phi}_{e}=\sum\nolimits_{i}\frac{\delta \Phi _{e}^{(i)}}{\delta t}%
=\sum\nolimits_{i}\frac{M_{iT}^{2}}{\mathcal{L}_{i}+L_{i}}\dot{I}_{L}
\end{equation}%
and Eq. (\ref{Fi_e}) is obtained with the inductance $\widetilde{L}$, which
describes the response of the quantum system to the tank circuit signal,
given by:%
\begin{equation}
\widetilde{L}=\sum\nolimits_{i}\frac{M_{iT}^{2}}{\mathcal{L}_{i}+L_{i}}.
\label{L_tilda}
\end{equation}

Thus, we have obtained the system of equations (\ref{Leff}), (\ref{eq_for_V}%
), (\ref{inductance}), and (\ref{L_tilda}), which describe the interaction
of the classical tank circuit and the quantum circuits (qubits). More
accurate (quantum-mechanical) analysis would start the description from the
Hamiltonian of the whole system in terms of the operators for both the tank
circuit and the qubits with averaging the equations afterwards, as in Ref.
\cite{Greenberg02b} (see also the discussion about similar systems in \cite%
{Everitt05} and \cite{Zorin}). Since this analysis would yield the same
equations for the observable values ($V_{T}=\left\langle \widehat{V}%
_{T}\right\rangle $, etc.), we do not consider this procedure here in detail.

\section{Analysis of the response of the tank circuit}

The measurement consists in biasing the tank circuit with the current $%
I_{bias}=I_{A}\cos \omega _{rf}t$ and measuring both the phase shift $\alpha
$ and amplitude $V_{A}$ of the voltage $V_{T}=V_{A}\cos (\omega
_{rf}t+\alpha )$. (The oscillations can be considered close to the harmonic
form due to the small losses in the high-quality tank circuit which is
weakly coupled to the nonlinear qubits' inductances, see (\ref{Q}) and (\ref%
{k}) below.) Substituting these expressions for $I_{bias}$ and $V_{T}$ in
Eq. (\ref{eq_for_V}) and equating coefficients before $\sin \omega _{rf}t$
and $\cos \omega _{rf}t$, we obtain:%
\begin{eqnarray}
\tan \alpha  &=&\frac{R_{T}}{\omega _{rf}}\left( L_{eff}^{-1}-\frac{\omega
_{rf}^{2}}{\omega _{T}^{2}}L_{T}^{-1}\right) ,  \label{alpha} \\
V_{A} &=&R_{T}I_{A}\cos \alpha .  \label{VA}
\end{eqnarray}%
Expression for the phase shift is simplified for the tank circuit driven at
resonance, $\omega _{rf}=\omega _{T}=\left( L_{T}C_{T}\right) ^{-1/2}$:%
\begin{eqnarray}
\tan \alpha  &=&Q\frac{\widetilde{L}}{L_{T}-\widetilde{L}}\approx Q\frac{%
\widetilde{L}}{L_{T}}=Q\sum\nolimits_{i}k_{i}^{2}\frac{L_{i}\mathcal{L}_{i}^{-1}}{%
1+L_{i}\mathcal{L}_{i}^{-1}}\approx   \notag \\
&\approx &Q\sum\nolimits_{i}k_{i}^{2}L_{i}\mathcal{L}_{i}^{-1}.  \label{tan}
\end{eqnarray}%
Here it was assumed that $\widetilde{L}\ll L_{T}$ and $L_{i}\ll \mathcal{L}_{i}$; the
latter inequality assumes that qubits' loops have small inductances, the
former inequality is justified for large quality factor and small tank
circuit-qubit coupling constants:
\begin{eqnarray}
Q &=&\omega _{T}R_{T}C_{T}\gg 1,  \label{Q} \\
k_{i}^{2} &=&\frac{M_{iT}^{2}}{L_{i}L_{T}}\ll 1,  \label{k}
\end{eqnarray}%
and also $k_{i}^{2}Q\lesssim 1$ is assumed.

Alternatively to measuring the phase shift $\alpha $, Eq. (\ref{tan}), the
qubits' parametric inductances can be probed by measuring the amplitude $%
V_{A}$ of the voltage, which is optimal at the frequency $\widetilde{\omega }$
defined by the relation $\left( \omega _{T}-\widetilde{\omega }\right)
/\omega _{T}=(2Q)^{-1}$, that is at $\widetilde{\omega }=\omega _{T}\left(
1-(2Q)^{-1}\right) $, then from Eqs. (\ref{alpha})-(\ref{Q}) it follows:%
\begin{equation}
\left. V_{A}\right\vert _{\omega _{rf}=\widetilde{\omega }}\approx R_{T}I_{A}%
\left[ 1+\left( 1+Q\frac{\widetilde{L}}{L_{T}}\right) ^{2}\right] ^{-1/2}.
\end{equation}%
This relation can be rewritten, taking into account Eq. (\ref{tan}) and
assuming $\tan \alpha \lesssim 1$ (which is usually the case in experiment
\cite{Ilichev04}) in the form:%
\begin{equation}
\left. V_{A}\right\vert _{\omega _{rf}=\widetilde{\omega }}\approx \frac{1}{%
\sqrt{2}}R_{T}I_{A}\left( 1+\frac{1}{2}\left. \tan \alpha \right\vert
_{\omega _{rf}=\omega _{T}}\right) ,  \label{relation}
\end{equation}%
which shows the equivalence of the measurements via the amplitude and the
phase shift of the tank circuit voltage. Since in practice it is more
convenient to probe flux qubits via the phase shift \cite{Ilichev04}, \cite%
{Iligrinb}, we will consider in what follows the phase shift only.

For the case of small geometrical inductances $L_{i}$, we can neglect the
shielding current, then $\Phi ^{(i)}\approx \Phi _{x}^{(i)}$; we also assume
the weak coupling of the qubits to the tank circuit (Eq. (\ref{k})) and
neglect the first term in Eq. (\ref{Fi_x}), and define $\Phi _{dc}^{(i)}$ as
the constant part of the flux through $i$-th qubit (in practice it changes
adiabatically slow); hence $\delta \Phi ^{(i)}\approx \delta \Phi
_{x}^{(i)}\approx \delta \Phi _{dc}^{(i)}$ and for calculations Eq. (\ref%
{tan}) is supplemented by the relation:%
\begin{equation}
\mathcal{L}_{i}^{-1}\approx \frac{\partial I_{qb}^{(i)}}{\partial \Phi _{dc}^{(i)}}.
\label{inverse_induct}
\end{equation}%
Note that this value (namely, the r.h.s. of Eq. (\ref{inverse_induct})) can
also be interpreted as the magnetic susceptibility \cite{Smirnov}, \cite%
{Everitt05}.

\section{Inductance of superconducting qubits}

Consider the case of a single superconducting qubit in more detail. The
phase shift $\alpha $ probes the current in the qubit as following (we
rewrite the equations derived above):%
\begin{eqnarray}
\tan \alpha  &\approx &k^{2}QL\mathcal{L}^{-1},  \label{for_1qb} \\
\mathcal{L}^{-1} &\approx &\frac{\partial I_{qb}}{\partial \Phi _{dc}}, \\
I_{qb} &=&\left\langle \widehat{I}\right\rangle =Sp\left( \widehat{\rho }%
\widehat{I}\right) .  \label{Iqb}
\end{eqnarray}%
The latter equation can be rewritten for both phase-biased charge qubit \cite%
{Zorin02}, \cite{Krech02} and flux qubit \cite{Mooij}, taking into account
that $\widehat{I}=I_{circ}\widehat{\sigma }_{z}$, as following: $%
I_{qb}=I_{circ}\left\langle \widehat{\sigma }_{z}\right\rangle $ (here $%
\widehat{\sigma }_{z}$ is the Pauli matrix). For a phase-biased charge qubit
\cite{multiphoton} the circulating current $I_{circ}=I_{0}$ is phase
dependent and Eqs. (\ref{for_1qb}-\ref{Iqb}) show that there are two terms
contributing in the tank circuit's phase shift:%
\begin{equation}
\tan \alpha \approx k^{2}QL\left( \frac{\partial I_{0}}{\partial \Phi _{dc}}%
Z+I_{0}\frac{\partial Z}{\partial \Phi _{dc}}\right) ,  \label{ch_qb}
\end{equation}%
where $Z=\left\langle \widehat{\sigma }_{z}\right\rangle $ is the difference
between the ground and excited state populations. In a classical system or
in the ground state the difference between the energy level's populations is
constant, $Z=const$, and the second term in Eq. (\ref{ch_qb}) is zero. In
contrast, for the quantum system the interplay of the two terms is
essential, which was studied in Ref. \cite{multiphoton}. At this point it is
worthwhile to notice that the second term can dominate at resonant
excitation, as it was the case in the work \cite{multiphoton} (\textit{cf.}
Figs. 3 and 5 in \cite{multiphoton}). Hence in some cases this may be the
advantage of the impedance measurement technique. Another advantage of the
technique may be the possibility of the non-destructive measurement (see in
Refs. \cite{Greenberg02b}, \cite{Smirnov}, \cite{Lupascu}).

Consider now the case of a flux qubit in detail. The current operator is
defined in the flux basis \cite{Mooij}, $\widehat{I}=I_{P}\widehat{\sigma }%
_{z}$, where $I_{P}$ stands for the amplitude value of the persistent
current, and hence the value $\left\langle \widehat{\sigma }%
_{z}\right\rangle $ defines the difference between the probabilities of the
clockwise and counter-clockwise current directions in the loop: $%
\left\langle \widehat{\sigma }_{z}\right\rangle =P_{\downarrow }-P_{\uparrow
}=2P_{\downarrow }-1$. Then with Eqs. (\ref{for_1qb}-\ref{Iqb}) we obtain%
\begin{equation}
\tan \alpha \approx k^{2}Q\frac{LI_{P}}{\Phi _{0}}2\frac{\partial
P_{\downarrow }}{\partial f_{dc}},  \label{flux_qb1}
\end{equation}%
where $f_{dc}=\Phi _{dc}/\Phi _{0}-1/2$.

For calculation of the density matrix $\widehat{\rho }$ the Bloch equation
is conveniently used (see \textit{e.g.} Ref. \cite{ShKOK}). This equation
includes the relaxation and correspondingly is written in the energy basis.
Thus, we rewrite Eq. (\ref{flux_qb1}) after introducing the density matrix
in the energy representation in terms of the unity matrix $\widehat{1}$\ and
the Pauli matrices $\widehat{\tau }_{i}$: $\widehat{\rho }=\left( 1/2\right)
\left( \widehat{1}+X\widehat{\tau }_{x}+Y\widehat{\tau }_{y}+Z\widehat{\tau }%
_{z}\right) $ (\textit{i.e.} $Z$ is again the difference between the ground
and excited state populations) and obtain:%
\begin{equation}
\tan \alpha \approx k^{2}Q\frac{LI_{P}}{\Phi _{0}}\frac{\partial }{\partial
f_{dc}}\left( \frac{2\Delta }{\Delta E}X-\frac{2I_{P}\Phi _{0}f_{dc}}{\Delta
E}Z\right) .  \label{tan(a)_full}
\end{equation}%
Here $\Delta E=2\sqrt{\Delta ^{2}+(I_{P}\Phi _{0}f_{dc})^{2}}$ is the
distance between the stationary energy levels and $\Delta $ is the tunneling
amplitude; about the details of this transition from current representation
to energy representation see in Ref. \cite{KOSh07}.

For one flux qubit in the ground state ($X=0$, $Z=1$) it results in the
following:

\begin{equation}
\tan \alpha \approx -k^{2}QL\frac{\Delta ^{2}I_{P}^{2}}{\left( \Delta
E/2\right) ^{3}}.  \label{alfagr}
\end{equation}%
It is important to notice that we obtained the result for the ground state,
Eq. (\ref{alfagr}), which coincides with the earlier obtained results (see
Eqs. (3-4) in \cite{Ilichev04}), but in different way -- by differentiating
the probability $P_{\downarrow }$, Eqs. (\ref{flux_qb1}-\ref{tan(a)_full}).
For the description of the flux qubit in the thermal equilibrium one has to
put $X=0$ and $Z=\tanh \left( \Delta E/2T\right) $ in Eq. (\ref{tan(a)_full}%
); then by plotting the phase shift versus the magnetic flux, $f_{dc}$, for
different temperatures, one can obtain the suppression and widening of the
zero-bias dip (that is in the vicinity of $f_{dc}=0$) as it was observed in
the experiment presented in Ref. \cite{1qb} in Fig. 3 (a), which is one more
confirmation of our consideration. For example, the zero-bias dip (that is $%
\tan \alpha $ at $f_{dc}=0$) is described by the r.h.s. of Eq. (\ref{alfagr}%
) multiplied by the factor $\tanh \left( \Delta /T\right) $.

If the first term in the bracket in Eq. (\ref{tan(a)_full}) can be neglected
(which in concrete case should be checked, but this is usually valid for
small driving amplitude $\Phi _{ac}$, see \textit{e.g.} in \cite{CohTan}),
then the expression is simplified:

\begin{eqnarray}
\tan \alpha &\approx &-k^{2}Q\frac{LI_{P}}{\Phi _{0}}\left[ \frac{\partial }{%
\partial f_{dc}}\left( \frac{2I_{P}\Phi _{0}f_{dc}}{\Delta E}\right) \cdot
Z\right. +  \notag \\
&&+\left. \frac{2I_{P}\Phi _{0}f_{dc}}{\Delta E}\cdot \frac{\partial Z}{%
\partial f_{dc}}\right] .  \label{approx}
\end{eqnarray}%
Note that at $f_{dc}=0$: $\alpha \sim Z$, which means that $\alpha $ probes
the changes of the upper level population.\

If a qubit is resonantly excited with the driving frequency $\omega $, then
the partial energy levels occupation probability $Z$ has the
Lorentzian-shape dependence on $f_{dc}$. It follows that the derivative $%
\partial Z/\partial f_{dc}$ takes the shape of a hyperbolic-like structure,
\textit{i.e.} it changes from a peak to a dip in the point of the resonance
at $\Delta E(f_{dc})=\hbar \omega $.

\section{Conclusion}

The impedance measurement technique for the tank circuit being coupled to
the system of qubits was studied. The tank circuit was considered to be
driven by the rf current and the voltage $V_{T}$ to be measured. The main
results of the work concern the phase shift $\alpha $ and the amplitude $%
V_{A}$ of the voltage $V_{T}$. It was obtained how the phase shift $\alpha $
is related to the parametric inductances of the qubits $\mathcal{L}_{i}$, Eq. (\ref%
{tan}).\ It was shown that the dynamics of the qubits can be studied via the
amplitude as well as via the phase shift, Eq. (\ref{relation}). The
derivations of these equations were presented in detail in order, first, to
make all the assumptions clear (small loops' inductances $L_{i}$, weak
driving of the tank circuit $I_{bias}$, high quality factor $Q$ and small
couplings $k_{i}$, slow dynamics of the tank circuit in comparison with
qubits, $\omega _{T}\ll \omega ,\Delta E$), and, second, to show how the
parametric inductances of the qubits should be defined, by introducing the
difference between the energy levels occupation probabilities, $Z$. We
obtained that the expression for the phase shift $\alpha $ in general
contains both terms proportional to $Z$ and proportional to $\partial
Z/\partial f_{dc}$, Eqs. (\ref{ch_qb}) and (\ref{approx}). If the latter
term dominates the resonant excitations are visualized as hyperbolic-like
structures on the dependence of the phase shift $\alpha $ on the dc flux $%
f_{dc}$.

I would like to thank A.N. Omelyanchouk and E. Il'ichev for stimulating
discussions and V.I. Shnyrkov, Ya.S. Greenberg, and W. Krech for helpful
comments. The work was supported by INTAS under the Fellowship Grant for
Young Scientists (No. 05-109-4479). The hospitality of \ Institute for
Physical High Technology (Jena, Germany) is acknowledged.

\end{document}